\documentclass[aps,prl,reprint,superscriptaddress,amsmath,amssymb]{revtex4-1}

\usepackage[draft]{changes}
\usepackage{graphicx}
\usepackage{subfigure}
\usepackage{import}
\usepackage{color}
\usepackage{amsmath, amssymb}
\usepackage{transparent}
\usepackage{braket}
\usepackage{soul}
\usepackage{dcolumn}
\usepackage[amssymb,textstyle,thinspace,thinqspace]{SIunits}
\usepackage{lmodern}
\usepackage{epstopdf} 

\begin{document}


\title{Landau polaritons in highly non-parabolic 2D gases in the ultra-strong coupling regime}


\author{Janine Keller}
\email[]{janine.keller@phys.ethz.ch}
\affiliation{Institute for Quantum Electronics, ETH Z\"{u}rich, 8093 Z\"{u}rich, Switzerland}

\author{Giacomo Scalari}
\email[]{scalari@phys.ethz.ch}
\affiliation{Institute for Quantum Electronics, ETH Z\"{u}rich, 8093 Z\"{u}rich, Switzerland}

\author{Felice Appugliese}
\affiliation{Institute for Quantum Electronics, ETH Z\"{u}rich, 8093 Z\"{u}rich, Switzerland}

\author{Shima Rajabali}
\affiliation{Institute for Quantum Electronics, ETH Z\"{u}rich, 8093 Z\"{u}rich, Switzerland}

\author{Mattias Beck}
\affiliation{Institute for Quantum Electronics, ETH Z\"{u}rich, 8093 Z\"{u}rich, Switzerland}

\author{Johannes Haase}
\affiliation{Paul Scherrer Institute, 5232 Villigen, Switzerland}

\author{Christian A. Lehner}
\affiliation{Laboratory for Solid State Physics, ETH Z\"{u}rich, 8093 Z\"{u}rich, Switzerland}

\author{Werner Wegscheider}
\affiliation{Laboratory for Solid State Physics, ETH Z\"{u}rich, 8093 Z\"{u}rich, Switzerland}

\author{Michele Failla}
\affiliation{University of Warwick, Coventry, CV4 7AL, United Kingdom}

\author{Maksym Myronov}
\affiliation{University of Warwick, Coventry, CV4 7AL, United Kingdom}

\author{David R. Leadley}
\affiliation{University of Warwick, Coventry, CV4 7AL, United Kingdom}

\author{James Lloyd-Hughes}
\affiliation{University of Warwick, Coventry, CV4 7AL, United Kingdom}

\author{Pierre Nataf}
\affiliation{Institute for Theoretical Physics, ETH Z\"{u}rich, 8093 Z\"{u}rich, Switzerland}

\author{J\'er\^ome Faist}
\affiliation{Institute for Quantum Electronics, ETH Z\"{u}rich, 8093 Z\"{u}rich, Switzerland}


\date{\today}

\begin{abstract}
We probe ultra-strong light matter coupling between metallic terahertz metasurfaces and Landau-level transitions in high mobility 2D electron and hole gases. We utilize heavy-hole cyclotron resonances in strained Ge and electron cyclotron resonances in InSb quantum wells, both within highly non-parabolic bands, and compare our results to well known parabolic AlGaAs/GaAs quantum well (QW) systems. Tuning the coupling strength of the system by two methods, lithographically and by optical pumping, we observe a novel behavior clearly deviating from the standard Hopfield model previously verified in cavity quantum electrodynamics: an opening of a lower polaritonic gap.

\end{abstract}


\maketitle


\section{I. Introduction}

Light-matter interaction phenomena are the driving force of quantum optics, and have been investigated in platforms ranging from atoms \cite{RaimondRMP2001} to solid state systems \cite{devoret2007}. The creation of quasi-particles called cavity polaritons in solid-state-based systems enables strong, non-linear, photon-photon interactions. This allowed the observation of Bose-Einstein condensation in solids \cite{KasprzakNature2006}, superfluidity, quantized vortices and dark solitons, forming the fascinating field of  quantum fluids of light \cite{carusottoRevModPhys85299}. A cavity polariton exists when the light-matter coupling strength is larger than the dephasing rates of the individual components, which are then in \textit{strong coupling} with reversible energy exchange between light and matter. The vacuum Rabi frequency $\Omega_R$ quantifies this coupling strength and by tuning the magnitude of $\Omega_R$, different physical regimes can be explored. An increase of the interaction strength towards the transition frequency $\omega_{ij}$ leads to the so-called ultra-strong coupling regime \cite{Ciuti2005,FornDiaz:2018vs,Kockum:2019}. Usually negligible terms, such as the polarization self-interaction and counter-rotating terms, then have to be included in the Hamiltonian representing the system. This impacts the physics of the resulting quasiparticles: the ground state of an ultra-strongly coupled system contains virtual photons \cite{Ciuti2005, Hagenmueller2010}, with a number proportional to the normalized light-matter coupling ratio $\Omega_R/\omega_{ij}$ \cite{Ciuti:PRA:06:033811}. 


One way to enhance  $\Omega_R$ and thus drive the system into the ultra-strong coupling regime is to exploit  the collective enhancement due to the simultaneous coupling of $N$ material excitations to the same cavity mode, yielding an increased Rabi frequency $\sqrt{N}\,\Omega_R$.
The collective radiative coupling of material excitations in a reduced volume ($V < \lambda^3$) is also the basis of the phenomenon of Dicke superradiance and of the superradiant phase transition \cite{Dicke1954,Hepp1973,Wang1973}. Dicke superradiance has been observed in atomic systems \cite{Skribanowitz1973} and more recently the superradiant phase transition has been realized in a driven-dissipative system of cold atoms  \cite{Baumann2010}. Such phenomena have attracted great attention lately also in the solid-state community, with observation of superradiant-related physics in different experimental platforms \cite{Cong2016}, for example semiconductor quantum dots \cite{Scheibner2007} or quantum wells \cite{PhysRevLettLaurent2015}.

In cavity quantum electrodynamics (QED) within the dipolar approximation, in solid-state systems and in semiconductor intersubband systems such a phase transition is prevented by the so-called `no-go' theorem \cite{Bialynicki1979,Nataf2010,Todorov2012}. The term containing the squared vector potential ($A^2$ term, also called the diamagnetic term) in the minimal-coupling Hamiltonian shifts the energy dispersion towards higher energies, such that the lower branch can never become gapless, and accordingly can never reach the ground state and exhibit the critical point associated to the  Dicke quantum phase transition \cite{Emary2003}. 
Recently, ultra-strong coupling has been obtained with the so-called Landau polaritons  \cite{Scalari2012}, where the cyclotron transition of a semiconductor-based system is optically coupled to a cavity in the THz range. Using meta material-based resonators  \cite{Scalari2012,Maissen2014} or Fabry-P\'erot cavities \cite{Li2018}  it has been possible to explicitly distinguish the $A^2$-term from the vacuum Bloch-Siegert shift related to the counter-rotating terms of the interaction.Suggestions to circumvent the no-go theorem in cavity QED include the use of multi-level atomic systems \cite{Baksic2013} or by considering systems with linear dispersion relation, like in graphene, although this question is still under debate \cite{Hagenmuller2012,Chirolli2012}. Other works showed, that also by including Coulomb dipole-dipole interaction \cite{Keeling2007,Vukics2014,DeBernardis2018} one could possibly restore a Dicke-like model. Other investigations by D. De Bernardis et al. \cite {PhysRevA.98.053819} deal with the breakdown of the gauge invariance, which is also subject of other recent investigations \cite{Malekakhlagh2017,DiStefano2018}. Very recently, P. Nataf et al. \cite{Nataf_PhysRevLett2019} included the Rashba spin-orbit interaction under a static magnetic field and found a magnetostatic instability which opens a promising way towards Dicke superradiant phases.

Depending on the shape of the confining potential of an intersubband transition, an additional higher order resonance which reduces the oscillator strength of the main mode, could lead to renormalized energies in the dipolar gauge (in contrast to the Coulomb gauge) but not to a Dicke phase transition \cite{PhysRevA.98.053819}. 


In experiments on Landau polaritons with two dimensional electron gases (2DEGs) in AlGaAs/GaAs QWs, which exhibit parabolic in-plane band dispersion, a coupling ratio beyond unity \cite{Bayer2017} has been achieved and a very good agreement of the polaritonic dispersion with the Hopfield-like Hamiltonian \cite{Hagenmueller2010} including all counter-rotating and diamagnetic terms has been verified \cite{Bayer2017,Scalari2012,Scalari2013,Scalari2014,Maissen2014,Maissen2017,Geiser2012,Zhang2016}. It has to be noted that the model developed in Ref.\cite{Hagenmueller2010} that was adopted in the previously cited papers and in this work has been originally developed for Fabry-P\'erot resonators. Here, striving to engineer ultra-strong coupling deviating from a standard Hopfield model in a purely ground state system, we utilize two QW systems with a non-parabolic 2D electron or hole gas: a strained germanium quantum well (s-Ge QW) with an occupied non-parabolic heavy-hole band and indium antimonide quantum wells (InSb QW) with extremely light mass electrons. We couple the 2DHG/2DEG Landau-level transitions ($\omega_{ij} = \omega_{cyc}$) to a terahertz (THz) metamaterial resonator with subwavelength electric field confinement. In the present work with s-Ge and InSb QWs, we systematically scale our cavity frequency $f_{LC} = \omega_{LC}/2\pi$ lithographically to control the coupling rate, and observe an opening of a  polaritonic gap below the cold cavity frequency.  The energy of the lower polariton branch no longer reaches the cold cavity frequency at large detunings  (high magnetic fields) as in the Hopfield model \cite{Hagenmueller2010}. A similar effect is observed  when optically pumping an intentionally undoped s-Ge QW creating more carriers in the QW.  We modify the standard Hopfield Hamiltonian including an \textit{effective} reduction of the diamagnetic terms to capture the purely experimentally observed change, obtaining a very good agreement to our experimental polariton branches.  

\section{II. Non-parabolic quantum wells}

\begin{figure}
\includegraphics[width=\columnwidth]{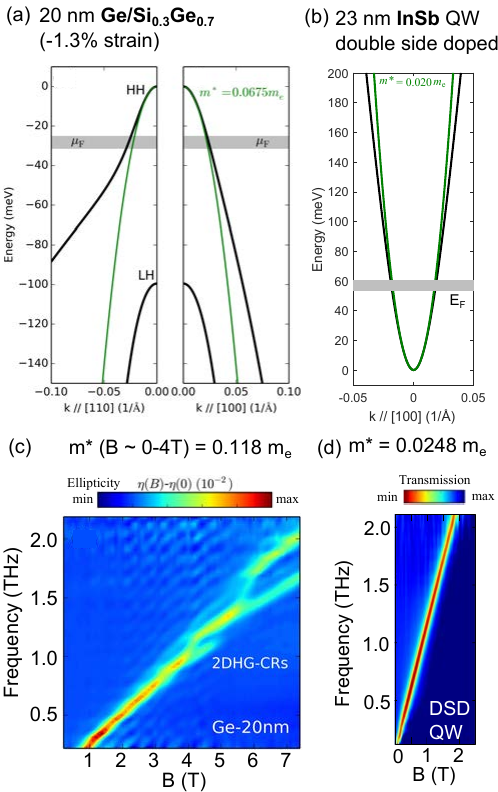}%
\caption{Calculated in-plane bandstructure of (a) the heavy hole (HH) and light hole (LH) band of the s-Ge QW and (b) the conduction band of an InSb QW (double side doped (SSD)). The shaded gray area indicates the chemical potential $\mu_F$ / Fermi energy $E_F$ of each sample. The measured cyclotron resonance transmission spectra are displayed as function of magnetic field for the (c) s-Ge QW (from Ref. \cite{Failla2016}, expressed in units of ellipticity $\eta$) and (d) InSb QW.}%
\label{fig:1}%
\end{figure}

The matter part of our coupled system, the inter-Landau level transition, is tunable in energy by an external static magnetic field as the cyclotron resonance scales as $\omega_{cyc} = eB/m^*$. The cyclotron resonance is directly accessible in transmission THz time domain spectroscopy (as shown in Fig. \ref{fig:1} (c), (d)), and the effective mass of the carriers can be deduced by a linear fit to the measured resonance. In contrast to a standard and well known AlGaAs/GaAs QW, the s-Ge QW and InSb QWs exhibit additional and more complex properties, appealing to conduct ultra-strong coupling experiments, which we compare then to the standard AlGaAs/GaAs QW. The s-Ge and InSb QWs are showing e.g. heavy non-parabolicity, strain and spin-orbit interaction \cite{Myronov2014,Failla2015,Failla2016,Lehner2018,Khodaparast2004}, with a heavier and lighter cyclotron effective mass than the standard AlGaAs/GaAs quantum wells, for which the cyclotron mass, due to the electron confinement, is found to be $m^*_{GaAs} = 0.071\,m_e$ ~\cite{Maissen2014}.\\
The s-Ge QW has a thickness $L = 20$\,nm, heavy-hole (HH) density  $1.3\times10^{12}$\,cm$^{-2}$, effective mass $m^*_{HH} = 0.118\,m_e$, $g$-factor $5.0$, and mobility $1.5\times10^6$\,cm$^2$V$^{-1}$s$^{-1}$ \cite{Failla2016}. In this system the 1.3\,\% biaxial compressive strain, provided by the $\text{Si}_{0.3}\text{Ge}_{0.7}$ barriers, lifts the degeneracy of the heavy-hole and light-hole band at the $\Gamma$-point. The band structure of this structure was calculated using the $6\times6$ $k \cdot p$ method following Ref.\ \cite{Sun2007}, including the heavy-hole, light-hole and split-off bands, and a wavevector $k_z = \pi/L$ to include the influence of quantum confinement. As shown in Fig. \ref{fig:1}(a), the heavy-hole band is strongly non-parabolic due to the applied strain. The effective mass at the $\Gamma$-point is $m^*_{HH} = 0.0675\,m_e$, as obtained from a parabolic fit at low wavevectors (green line). The gray shaded area indicates the range of the oscillatory chemical potential $\mu_F$, which lies clearly in the region that is no longer within the parabolic approximation. Note that within this range the in-plane dispersion remains essentially isotropic.

The InSb QWs have a strongly non-parabolic conduction band with a very small bandgap ($\approx 180\,$meV \cite{Adachi1989}), featuring a very light effective mass electron which we determined by cyclotron resonance measurements to be $m^*_e = 0.0248\,m_e$ for a double side doped (DSD) quantum well and $m^*_e = 0.0243\,m_e$ for a single side doped (SSD) QW. The effective mass of the InSb DSD QW at the $\Gamma$-point is $m^*_e = 0.020\,m_e$, as obtained from a parabolic fit at low wavevectors (Fig. \ref{fig:1} (b) green line). The DSD QW has a thickness of $L = 23$\,nm, an electron density of $4.9\times10^{11}$\,cm$^{-2}$ with a mobility of $3.49\times10^5$\,cm$^2$V$^{-1}$s$^{-1}$, and the SSD QW has a thickness of $L = 21$\,nm, an electron density of $3.65\times10^{11}$\,cm$^{-2}$ with a mobility of $2.03\times10^5$\,cm$^2$V$^{-1}$s$^{-1}$. Details on the growth of such QWs are published in Ref. \cite{Lehner2018}.

\section{III. Lithographic tuning}

\begin{figure}
\includegraphics[width=0.8\columnwidth]{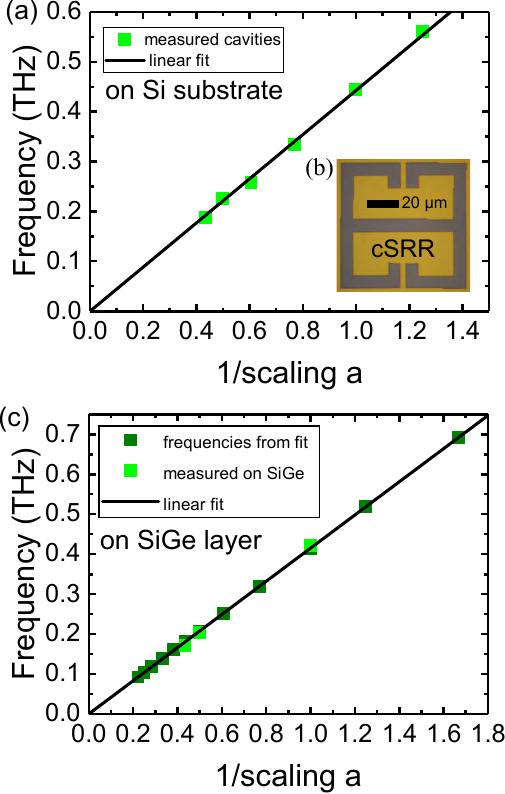}%
\caption{(a) Scaling of the cavity frequency on a Si substrate. In bright green squares are the measured frequencies with a linear fit (black line). In the inset (b) the cSRR is displayed for scaling $a = 2$. In (c) the scaling on a $3\,\micro\meter$ layer of SiGe on Si is displayed with the measured frequencies in bright green squares from which with a linear fit (black line) the used frequencies in dark green squares are deduced.}%
\label{fig:2}%
\end{figure}

For the cavity we chose complementary split ring resonators (cSRRs), which can be described by a lumped element electric circuit model with a characteristic LC-resonance where the vacuum electric field fluctuations are greatly enhanced due to the strongly sub-wavelength cavity volume \cite{Scalari2012,Scalari2014}. We design a cavity with a LC-resonance $f_{LC} = \omega_{LC}/2\pi$ and then scale the geometry of the resonator by a linear factor $a$ (from $a = 0.5$ to $a = 2.3/2.4$ on all QWs) with constant metal thickness (4\,nm Ti and 200\,nm Au). The lithographic tuning was experimentally verified and measured on a bare Si substrate and on GaAs in our previous work \cite{Maissen2017}, revealing a linear frequency scaling with the inverse of the geometrical scaling factor $f_{LC} \propto a^{-1}$, as shown in Fig. \ref{fig:2} (a),  with an optical micrograph of one cavity of the full array with scaling $a = 2$ shown in the inset (b).\\

As the frequency of the cavity depends on the dielectric environment in close vicinity, the cold cavity frequencies, i.e. the frequencies without  free carrier contributions, for the s-Ge and InSb QW samples were further determined by depositing three cavity arrays chosen from across the frequency range of $f \approx 200\,$GHz to $f \approx 900$\,GHz (= different scalings) on buffer layer structures. The buffer layers have the same growth structure as the QW sample, thus the same refractive index, but do not contain a QW. From a linear fit to the measured cavity frequencies (Fig. \ref{fig:2} (c), bright green) on the reference structures the expected bare cavity frequencies for all arrays of cSRR deposited on the QWs are deduced (see Fig. \ref{fig:2} (c), dark green rectangles). Additionally, the lithographic accuracy (electron beam lithography for s-Ge QW and photolithography for InSb and GaAs QWs) is verified using scanning electron microscopy and the frequencies of the expected cold cavity are corrected accordingly (this correction remained very small $< 2\% $).

\begin{figure}
\includegraphics[width=\columnwidth]{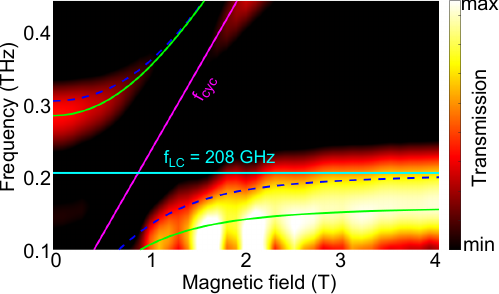}%
\caption{THz transmission of a resonator at $f_{LC} = \omega_{LC}/2\pi = 208\,\giga\hertz$ as a function of magnetic field. White areas indicate high transmission and the polariton branches. The cold cavity frequency $f_{LC} = \omega_{LC}/2\pi$ is shown by the solid cyan lines, and the solid magenta lines show the cyclotron frequency $f_{cyc} = \omega_{cyc}/2\pi$. Polariton dispersion fits are shown with the Hopfield model (dashed blue lines) and with a fitted reduced prefactor $d$ (solid green lines). }%
\label{fig:3}%
\end{figure}

Probing the coupled samples with THz time domain spectroscopy (see S.M. for details \cite{Supplementary}) in transmission, we measure the polariton dispersion of each frequency and resonator array on the s-Ge, InSb and GaAs QWs at a temperature of $3\,$K. 

One example of such measurement of a s-Ge QW at high filling factors, thus at low frequencies respectively, is shown in Fig. \ref{fig:3}. The bare cavity frequency for the shown scaling factor $a = 2$ is at $f_{LC} = 208$\,GHz (solid cyan line), which lies between the frequencies of the polariton branches at high and low magnetic field, with $f_{LP} = 165$\,GHz (lower polariton (LP)) and $f_{UP} = 292$\,GHz (upper polariton (UP)), respectively. This is a very striking and peculiar feature, as in the standard Hopfield-like Hamiltonian \cite{Ciuti2005,Hagenmueller2010} used to describe the ultra-strong coupling, one asymptotically recovers the cold cavity frequency at high magnetic fields.

\section{IV. Tuning by optical pumping}

\begin{figure}
\includegraphics[width=0.9\columnwidth]{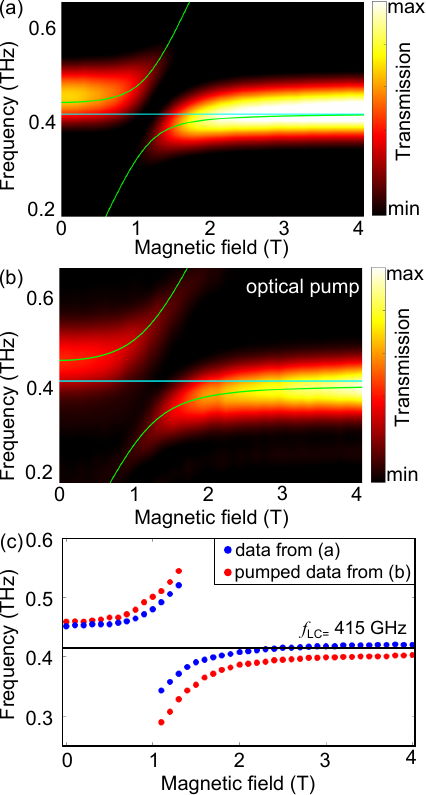}%
\caption{THz transmission spectra in of the s-Ge QW with a cavity frequency of $f = 415 $ GHz are displayed. In (a) without additional optical pump and (b) with optically pumping the system. In (c) the extracted intensity maxima of the polariton branches from the spectra in (a) and (b) are shown together for comparison.}%
\label{fig:4}%
\end{figure}

Additionally to the lithographic tuning, it is an attractive option to use an optical pump to manipulate \textit{in-situ} the polariton state. It has been shown that organic molecules can be photochemically changed reversibly by UV and VIS radiation to switch from the weak to ultra-strong coupling regime \cite{Schwartz2011}. Recently, exploiting the alignment of the light polarization in respect to carbon nano-tubes enabled such continuous tuning from weak and ultra-strong coupling\cite{GaoNatPhys2018nanotubes}. In the work of G. G\"unter et al., \cite{Gunter2009}, the authors were able to show an ultra-fast switch-on of ultra-strong coupling by optically exciting carriers in a QW to enable the intersubband transition. For the current study, we use an optical pump to change the carrier concentration in a s-Ge QW. 
As a pump we used part of our Ti:Sapphire beam at 800 nm with an optical fluence of approximately 7 $\mu$J/cm$^2$ (the beam area onto the sample was $\sim 3$ mm$^2$.)
 No time-delay-dependent carrier density can be observed (see Supplementary Material), suggesting a lifetime much longer than the time lapse between the laser pulses at our laser repetition rate of $80$ MHz. As a result, no special care had to be taken to adjust the delay between the optical pump and THz probe pulses: our optical pumping is in fact a way to change "statically" the carrier density of the sample employing the same cavity.  \\
Working with a s-Ge QW which has the same growth structure as the s-Ge layer used for the lithographic tuning but no intentional doping, we still observe a cyclotron resonance from the residual carriers in the QW with a mass of $m^* = 0.076\,m_0$. Working in a non-parabolic QW as s-Ge, the cyclotron mass is carrier-density-dependent, thus optically pumping the QW leads to an increase of the cyclotron mass to $m^* = 0.08\,m_0$ (see supplementary material for details). The employed cavity is the same as scaling $a =1$ from the lithographic tuning study with a frequency of $f = 415\,\tera\hertz$. THz transmission spectra as function of magnetic field are taken without additional optical pump in Fig. \ref{fig:4} (a) and with optical pump in Fig. \ref{fig:4} (b). The coupling strength increases from $\Omega/\omega = 0.17$ without pumping to $\Omega/\omega = 0.25$ with additional optical pump. In the Hopfield model, the increased coupling strength would predict an increase of the upper polariton branch frequency, leading to a larger upper polariton gap while far detuned the lower polariton branch still reaches the unchanged cavity frequency in the Hopfield model. In Fig. \ref{fig:4} (c) where the maxima of the polariton branches with and without optical pump are extracted and plotted together, it is  clearly visible that the lower polariton branch does not reach the cold cavity frequency far detuned, thus opening a lower polariton gap, as also observed in the measurements that use lithographically tuning.

\section{V. Analysis and discussion}

\begin{figure}
\includegraphics[width=0.95\columnwidth]{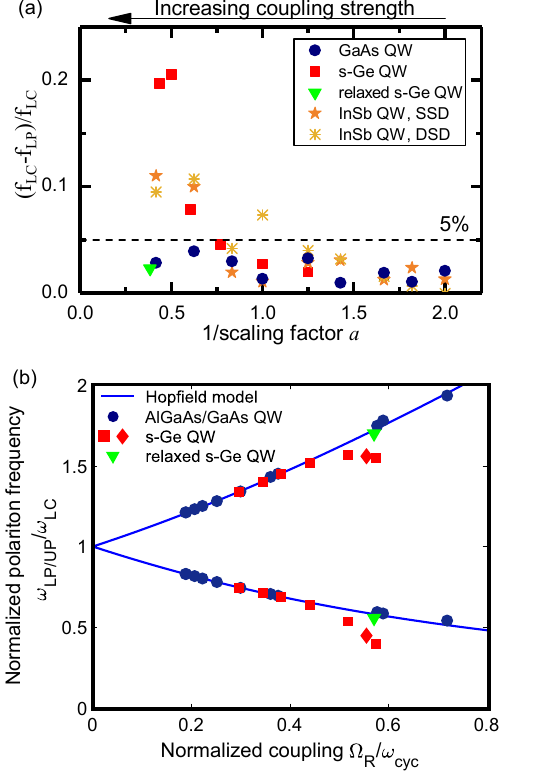}%
\caption{In (a) the deviation of the lower polariton  frequency $f_{LP}$ to the cold cavity frequency $f_{LC}$ is displayed for GaAs QW (dark blue circles), s-Ge QW (red squares) and InSb QWs (orange (single side doped QW) and yellow (double side doped QW) stars) as function of the geometrical scaling, where the coupling strength increases towards the left. (b) Normalized polariton frequencies $\omega_{LP/UP}/\omega_{LC}$ at the position of minimal splitting versus the normalized coupling strength $\Omega_R/\omega_{cyc}$. The theoretical Hopfield-model is displayed with blue solid lines. Experimental data points for parabolic 2D electrons in GaAs QWs fit the Hopfield model (dark blue circles \cite{Maissen2014, Maissen2017}). Experimental data points for the non-parabolic 2D heavy-holes in a s-Ge QW are displayed in red squares (series 1) and diamonds (series 2). The bright green triangles are for a partially relaxed s-Ge QW of sample series 2.}%
\label{fig:5}
\end{figure}

The deviation of the lower polariton frequency $f_{LP}$ at high magnetic fields from the cold cavity frequency $f_{LC}$ is extracted for all lithographically tuned cavities on GaAs, s-Ge and InSb QWs and displayed in Fig. \ref{fig:5} (a) as function of the inverse scaling factor \textit{a}. In the case of the GaAs QW (Fig. \ref{fig:5} (a), dark blue circles), we verify again that the deviation for all frequencies ($f \approx 200-900$\, GHz) is less than $5\%$ (and less than $15$ GHz in absolute terms). For the s-Ge and InSb QWs instead, we observe an increasing deviation with increasing coupling strength, up to $20\%$ for the s-Ge QW (Fig. \ref{fig:5} (a), red squares, $43$ GHz absolute deviation) and $10\%$ for the InSb QWs (Fig. \ref{fig:5} (a), yellow and orange stars). Interestingly, one cavity array ($f_{LC} = 160\,$GHz) of the s-Ge QWs (Fig. \ref{fig:5} (a), bright green triangle), which shows a characteristic crosshatch pattern of strain relaxation (see S. M. \cite{Supplementary}), has a polariton frequency of the lower branch very close to the cold cavity frequency again, with a deviation of less than $5\%$ as for the GaAs QW.\\

The Hopfield-like Hamiltonian can be written as the sum of different contributions \cite{Hagenmueller2010}: $H = H_{mat}+H_{int}+H_{dia}+H_{cavity} $,  with the material excitation $H_{mat}$, the bare cavity electromagnetic field $H_{cavity}$, the interaction term $H_{int}$  and the diamagnetic term $H_{dia}$ arising from the self-interaction of the light, including all counter-rotating terms. The diamagnetic term is \cite{Ciuti2005,Hagenmueller2010}
$H_{dia} = \hbar \sum_k{\left(D_k(a^\dagger_ka_k+a_ka^\dagger_k)+D_k(a_k a_{-k}+a^\dagger_k a^\dagger_{-k})\right)}$
and leads to a renormalization of the polariton energies. The value $D$ of the diamagnetic terms is in the case of parabolic dispersion approximated as $D_k = D \approx \Omega_R^2/\omega_{cyc}$ \cite{Hagenmueller2010} by evaluating the Thomas-Reiche-Kuhn sum rule (also known as the f-sum rule) \cite{Ciuti2005}. 

To capture the observed different renormalization of the polariton energies, we introduce here a parameter $d$ which we use to reduce effectively the strength of the diamagnetic term $D = d\,\Omega^2_R/\omega_{cyc}$. In the effective model, a polaritonic gap opening with respect to the bare cavity frequency for \textit{both} polariton branches is predicted (see details in S. M. \cite{Supplementary}), just as observed in Fig. \ref{fig:3}. 
We fit the measured polariton branches to extract the normalized coupling ratio $\Omega_R/\omega_{cyc}$ following the procedure described in Ref. \cite{Scalari2012}, but implementing the modified Hopfield model with the prefactor $d$ as an additional fitting parameter for the effective diamagnetic term $D = d \,\Omega^2_R/\omega_{cyc}$. The fit (solid green line) is in very good agreement with the measured polariton branch dispersion, yielding a prefactor $d = 0.7$ (normalized root mean square (RMS) deviation below $\approx 3\%$) for the shown cavity frequency. The normalized coupling rate is as large as $\Omega_R/\omega_{cyc} = 0.57$. If, instead, one keeps the prefactor at $d = 1$ for the standard parabolic Hopfield model, the fit (dashed blue line) does not describe the measured data very accurately (normalized RMS deviation rises to above $15\%$).\\

One efficient way to plot  \cite{Nataf2010} the polariton dispersion is to show the normalized polariton frequencies as function of normalized coupling strength, as in Fig. \ref{fig:5} (b). Thus, for each anti-crossing, which we measure as transmission spectra as function of magnetic field, we extract the upper and lower polariton frequency $\omega_{LP/UP}$ at the point of the minimal splitting of the two branches (= vacuum Rabi frequency) and normalize to the cold cavity frequency $\omega_{LC}$ (exact step by step procedure can be found in the S.M. \cite{Supplementary}). In the Hopfield model, the point of minimal splitting corresponds strictly to the resonant condition $\omega_{LC} = \omega_{cyc}$, whereas in an effective model with $d<1$, this point of minimally splitted branches shift towards lower magnetic fields (as for a pure Dicke model with $D \equiv 0$, see S. M. \cite{Supplementary}). The solid blue line in Fig. \ref{fig:5} (b) corresponds to the calculated standard Hopfield model, which agrees well with our previous experiments on AlGaAs/GaAs QWs \cite{Scalari2012,Scalari2013,Scalari2014,Maissen2014,Maissen2017}. Full dark blue circles show our data from Refs. \cite{Maissen2014,Maissen2017}, nicely agreeing with the Hopfield model. For the scaling study on s-Ge QWs (Fig. \ref{fig:5} (b), red squares and diamonds) instead, we observe a clear deviation of our measured polariton frequencies from the calculated Hopfield dispersion. The upper and the lower polariton frequencies are at lower frequencies than expected by the Hopfield model at high normalized couplings $\Omega_R/\omega_{cyc}$. The prefactor $d$ and the Rabi frequency serve as fitting parameters and it is especially notable, that the obtained value for $d$ is dependent on the coupling strength and does not represent a constant value in our fitting result (see also data tables in the S.M. \cite{Supplementary}). The deviation from the Hopfield model increases with increasing coupling strength, which we achieve by scaling the cavity frequency to lower values and thus corresponds to an anticrossing a lower magnetic fields. The measurement with largest deviation shown is the same as in Fig. \ref{fig:3} as described before with $d = 0.7$ for $\Omega_R/\omega_{cyc} = 0.57$. Further measurements with even lower cavity frequencies result in only the observation of the upper branch, as the lower polariton branch is lower than the experimentally observable frequency region (THz-TDS bandwidth $> 0.1$ THz). In the supplementary material, for these measurements where only the upper branch is measured experimentally, we also include in the same plot the frequency of the lower polariton branch that we extrapolate by assuming that the normalized light-matter coupling strength scales linearly with the size of the resonator   \cite{Supplementary}.\\

Additionally, in Fig. \ref{fig:5} (b) we also include a measurement at $f_{LC} = 160\,\giga\hertz$ with a partially relaxed s-Ge QW sample (as in Fig. \ref{fig:5} (a)), where the semiconductor material shows a crosshatch pattern, characteristic of strain relaxation (see S. M. \cite{Supplementary}). The best fit of the polariton branches yields a prefactor $d = 0.95$, with the normalized polariton frequencies (Fig. \ref{fig:5}, green triangles) close to the Hopfield model again. This result suggests that the strain plays a critical role for the observed renormalization of the polariton energies.

In conclusion, we report a solid-state system in the ultra-strong coupling regime that exhibits a mode softening of the polariton branches compared to the standard Hopfield model, where the lower polariton never reaches the ground state, regardless of how high the normalized coupling rate is. We found again that the Hopfield model\cite{Hagenmueller2010}, fits our data within the experimental errors for the GaAs/AlGaAs 2DEGs.  In contrast, in the s-Ge QW and InSb QW systems the polariton branches are indeed experimentally observed at lower frequencies. We capture this change by an effective model including a reduced diamagnetic term, which at high filling factor is about $30\%$ less than in the Hopfield model. Key features of our system, which might lead to a theoretical model to predict this observed change, include the strain in the systems, non-parabolic band dispersions and (Rashba)\cite{Nataf_PhysRevLett2019} spin-orbit coupling effects \cite{Myronov2014,Failla2015,Failla2016,Lehner2018,Khodaparast2004}. Which physical effect leads to the observed deviation from the Hopfield and whether this could lead to a Dicke quantum phase transition remains an open question and needs to be investigated in the future. However, with our results we clearly enter an uncharted regime of the ultra-strong coupling with a \textit{purely ground state, solid state system} that does not fall within  the validity of a Hopfield model. Furthermore, contributing effect of the very tight electric field confinement in the metamaterial resonator cannot be ruled out, as most theoretical models \cite{Hagenmueller2010} still consider Fabry-Perot type cavities with optical confinement over a wavelength of light. The strong electric field gradients that occur at deep subwavelength dimensions in our metamaterials will enhance, through Maxwell's equation, the vector potential and therefore magnetic coupling to the system.  Exploring the nature of the ground state and its excitations in this parameter region would be highly relevant and could answer fundamental questions concerning the possibility of a Dicke superradiant transition outside of driven systems \cite{Baumann2010}. Experimentally, this study would benefit from further increasing in the light-matter coupling strength by, for example, increasing the number of quantum wells, the hole density, or further down-scaling of the resonator frequencies. Investigating other complex solid state material system might also be able to help to disentangle the possible causes and shed light on the origins of the observed opening of a lower polariton gap and its implications.\\



We thank Elena Mavrona for supporting measurements and Curdin Maissen for discussions. We acknowledge financial support from the ERC grant 340975-MUSiC and the EPSRC (UK). We also acknowledge financial support from the Swiss National Science Foundation (SNF) through the National Centre of Competence in Research Quantum Science and Technology (NCCR QSIT) and Molecular Ultrafast Science and Technology (NCCR MUST).

%

\end{document}